%% file: main.tex
\documentclass[12pt]{article}
\usepackage{arxiv}

\include{preamble}

\mathtoolsset{showonlyrefs=true}

\title{Model Assessment for a Generalised Bayesian Structural Equation Model}

\author{
 Konstantinos Vamvourellis, Konstantinos Kalogeropoulos, Irini Moustaki\\
 Department of Statistics\\
  LSE\\
  \texttt{k.vamvourellis@lse.ac.uk}; \texttt{k.kalogeropoulos@lse.ac.uk};  \texttt{i.moustaki@lse.ac.uk} \\
}

\begin{document}
\maketitle

\input{abstract}


\input{sections}

\bibliography{references}

\appendix
\include{appendix}

\vfill\eject
\end{document}

%% file: preamble.tex
\usepackage{graphicx}

\usepackage{amsmath}
\usepackage{apacite}
\usepackage{booktabs}
\usepackage{mathtools}
\usepackage{appendix}

\newcommand{\bfu}{\mbox{\boldmath$\mathsf{u}$}}

\newcommand{\Ytil}{\widetilde{Y}}

\newcommand{\IW}{\mathcal{IW}}

\newcommand{\beq}{\begin{equation}}
\newcommand{\eeq}{\end{equation}}
\newcommand{\bea}{\begin{eqnarray}}
\newcommand{\eea}{\end{eqnarray}}
\newcommand{\ba}{\begin{array}}
\newcommand{\ea}{\end{array}}
\newcommand{\bi}{\begin{itemize}}
\newcommand{\ei}{\end{itemize}}
\newcommand{\ben}{\begin{enumerate}}
\newcommand{\een}{\end{enumerate}}
\newcommand{\bfY}{\mbox{\boldmath{$Y$}}}

\newcommand{\bfy}{\mbox{{\bf y}}}
\newcommand{\bfz}{\mbox{{\bf z}}}
\newcommand{\bfepsilon}{\bf \epsilon}
\newcommand{\bfe}{\mbox{\boldmath{$e$}}}
\newcommand{\bfeta}{\mbox{\boldmath{$\eta$}}}

%% file: abstract.tex
\begin{abstract}
The paper proposes a novel model assessment
paradigm aiming to address shortcoming of posterior predictive $p-$values, which provide the default metric of fit for Bayesian structural equation modelling (BSEM). The model framework of the paper focuses on the approximate zero approach, according to which parameters that would before set to zero (e.g. factor loadings) are now formulated to be approximate zero via informative priors \cite{MA12}. 
The introduced model assessment procedure monitors the out-of-sample predictive performance of the fitted model, and together with a list of guidelines we provide, one can investigate whether the hypothesised model is supported by the data. We incorporate scoring rules and cross-validation to supplement existing model assessment metrics for Bayesian SEM. The proposed tools can be applied to models for both categorical and continuous data. The modelling of categorical and non-normally distributed continuous data is facilitated with the introduction of an item-individual random effect that can also be used for outlier detection. We study the performance of the proposed methodology via simulations. The factor model for continuous and binary data is fitted to data on the `Big-5' personality scale and the Fagerstrom test for nicotine dependence respectively.

\end{abstract}

%% file: sections.tex
\section{Introduction}
Structural Equation Modelling (SEM) is a general framework for testing research hypotheses arising in psychology and social sciences in general \cite{Bollen1989}. Initial inference methods for SEM have mostly been frequentist, but recently their Bayesian counterpart has gained popularity \shortcite<see e.g.>{SHB99,DPB05,K14,MR15,VWR17}. 

In this paper, we focus on the Bayesian SEM framework introduced by \citeA{MA12}. Depending on the substantive theory to be tested, structural equation models impose restrictions on some  of the model  parameters. Usually, some parameters are set to zero and thus not estimated at all (e.g. cross-loadings, error correlations, regression coefficients). \shortciteA{MA12} suggested treating such parameters as approximate rather than exact zero, by assigning informative priors on them that place a large mass around zero; we will refer to this approach as the 
\emph{approximate zero framework}. The introduction of such informative priors is convenient in situations where there are concerns regarding the fit of the exact zero model. More specifically, it allows using the model as an exploratory tool to identify the source of model misfit. An alternative option for such a task is the use of modification indices \shortcite<e.g.>{MRN92}. More specifically, a modification index measures the improvement in model fit that would result if a previously omitted parameter were to be freely estimated. This can often lead to a model unsupported by the hypothesised substantive theory. Moreover, the greedy nature of the procedure does not guarantee convergence to an optimal model \cite{MA12,SMSD15,AMM15}. On the other hand, the approximate zero approach provides information on model modification in one go, while the hypothesised theory is reflected clearly via the priors on the loadings.  A related approach is to use Bayesian model searching, which could be implemented using spike and slab priors and stochastic search variable selection; see for example \shortciteA{LCL16}. 

In this paper, we aim to improve upon two aspects of the approximate zero framework. First, we note that the error correlations, the purpose of which is to allow for small deviations from the assumption of the items being independent given the latent factors, are only defined for normally distributed data. In order to generalise the approximate zeros framework so that it covers other distributions, e.g. binary data using the logit link, we introduce an item-individual random effect term in the measurement model. 
Second, we focus on the task of assessing model fit under the approximate zero framework \cite{VJ20,AM21a}. While the task of model fit assessment is challenging in its own right, things are further complicated in the case of the model specified under the approximate zero framework. More specifically, it is unclear whether fitting such models can result in good fit indices even when they are incorrectly specified \shortcite{SMSD15}. In terms of specific model fit indices, several approaches exist in the literature with the posterior predictive $p$-values (PPP) \cite{M94} being the most widely used. However, concerns have been raised regarding their suitability in this framework \shortcite<see e.g.>{HV18}. Special consideration has to be given to the choice of prior distributions for the model parameters, which can potentially affect the PPP performance \shortcite{MEC12,VMO18,L20}. Perhaps a more fundamental question is whether priors should be set on the basis of fit indices, rather than formal Bayesian model choice quantities, such as the Bayes factor. Nevertheless, the Bayes factor requires calculating the model evidence, or else marginal likelihood \shortcite{GSB17}, which can be quite a challenging task especially in models with latent variables \shortcite<e.g.>{LW04,VNM14}. Moreover, the Bayes factor is a relative measure and therefore does not directly address the question of whether a model fits the data well. \shortciteA{AMM15} suggest avoiding the use of the approximate zero model to reach binary decisions on goodness of fit, but instead use it as an exploratory tool leaving the choice up to the subject matter experts. 

Our approach aims towards reaching a middle ground between exploring lack of fit and assessing its severity. This is done by developing a decision framework that monitors the out-of-sample predictive performance to explore model misfit (i.e. validity of the hypothesised theory). The proposed decision framework uses collectively fit indices and scoring rules via cross-validation to examine whether the approximate zero parameters are picking up random noise rather than systematic patterns in the data. From a machine learning viewpoint, cross-validation is one of the standard tools to guard against overfit, while at the same time ensuring a good fit. The advantages of using cross-validation to measure model performance have been noted in the SEM context; see, for example, \shortciteA{MRN92}, \shortciteA{B00} and recommendation 4 of \citeA{SMSD15}. An intuitive argument in favour of cross-validation is that if a measurement scale does not generalise well in parts of the existing data it is highly unlikely that it will in future data. \shortciteA{MFR19} use the DIC and WAIC indices that can be viewed as approximate versions of cross-validation \shortcite{GHV14}, although their resulting approximation is not always satisfactory; see for example \citeA{P08}. From a Bayesian viewpoint, cross-validation, when combined with the log posterior predictive scoring rule, has tight connections with the model evidence and is less sensitive to priors \shortcite{FH20}. The proposed framework is developed by combining fit and out of sample predictive performance indices from different models. 

The paper is structured as follows. In Section 2, we define the generalised Bayesian SEM framework. Section 3 introduces the framework for assessing Bayesian SEM models. In Section 4, we illustrate and assess our methodology through several simulation experiments. Section 5 presents the analyses of two real applications: the first is a standard example on examining the `Big 5' personality factors on data from the 2005-06 British Household Panel Survey (BHPS), whereas the second example is on the Fagerstrom Test for Nicotine Dependence (FTND). Finally, Section 6 concludes with some relevant discussion and extensions.
The code for this work is available in the accompanying repository `bayes-sem' hosted on github\footnote{https://github.com/bayesways/bayes-sem/}.

\section{Generalised framework for Bayesian SEM}
\label{gbsem_sec:models}

\subsection{Model specification}
 Suppose there are $p$ observed variables (items) denoted by ${\bf y}=(y_1,\ldots,y_p)$ and that their associations are explained by $k$ continuous latent variables (factors) denoted by ${\bf z} = (z_1, \ldots, z_k)$. Categorical variables (binary and ordinal) can be accommodated in the same framework as for continuous data by assuming that the categorical responses are manifestations of underlying (latent) continuous variables denoted by ${\bf y^*} = (y^*_{1}, \ldots, y^*_{p})$. When continuous variables are analysed, $y_j=y^*_j$, ($j=1,\ldots,p$). The classical linear factor analysis model ({\it{measurement model}}) is: %
\begin{equation}
\label{gbsem_augmented}
{\bf y}^*_i = \alpha+\Lambda {\bf z}_i  + \bfepsilon_i, \;\; i=1,\ldots,n 
\end{equation}
where $\alpha$ is a $p\times 1$ vector of intercept parameters, $\Lambda$ is a $p\times k$ matrix of factor loadings and $n$ is the sample size. The vector of latent variables ${\bf z}_i$ has a Normal distribution, ${\bf z}_i \sim N_k(0, \Phi)$, where the covariance matrix $\Phi$ is either unstructured or defined by a parametric model that relates latent variables with each other and observed covariates ({\it structural model}). The $\bfepsilon_i$s are error terms assumed to be independent from each other and from the ${\bf z}_i$s. 

For binary data, the connection between the observed binary variable $y_j$ and the underlying variable $y^*_j$ is $y_{j}=\mathcal{I}(y_{j}^*>0)$. Similarly, for an ordinal variable with $m_j$ categories, $y_{j}= a $ if $ \tau_{a-1}^{(j)} < y^*_{j} \leq \tau_{a}^{(j)}, a=1,\ldots,m_j$ where $\tau_{0}^{(j)}=\infty, \tau_{1}^{(j)}<\tau_{2}^{(j)}<\ldots<t_{m_j-1}^{(j)}, \tau_{m_j}^{(j)}=+\infty$. 
More specifically, for a binary item $j$ and individual $i$, the probability of success (positive) response is given by:
\begin{equation}
P(y_{ij}=1 \mid {\bf z}_i)=P(y^*_{ij}>0 \mid {\bf z}_i)=P(\alpha_j+\Lambda_j {\bf z}_i  + \epsilon_{ij}>0\mid {\bf z}_i)=
P(\epsilon_{ij}<\alpha_j+\Lambda_j {\bf z}_i\mid {\bf z}_i)=F(\alpha_j+\Lambda_j {\bf z}_i),
\end{equation}
where $F$ stands for the cumulative distribution function (CDF) of $\epsilon_{ij}$. Finally, the model becomes:
\begin{equation}
F^{-1}\{P(y_{ij}=1)\}= \alpha_j+\Lambda_j {\bf z}_i,
\end{equation}
where $F^{-1}$ is the inverse of the CDF also known as the link between the probability of success and the linear predictor. Specific choices for the distribution of the error term $\bfepsilon_i$ lead to the following well known models:
\begin{equation}
\label{gbsem_augmented2}
\bfepsilon_i\sim\begin{cases} N(0_p, \Psi), \;\;\Psi=\text{diag}(\psi_1^2,\dots,\psi_p^2), & \text{if }{\bf y}_i\text{ is continuous}\\
N(0_p, \Psi), \;\;\Psi= I_p, &\text{if }{\bf y}_i \text{ is binary \& $F^{-1}$ is the inverse CDF of the normal}\\
 \prod_{j=1}^J\text{Logistic}(0,\pi^2/3), & \text{if }{\bf y}_i\text{ is binary \& $F^{-1}$ is the inverse CDF of the logistic},
\end{cases}
\end{equation}
where $0_p$ is a $p$-dimensional vector of zeros and $I_p$ denotes the identity matrix of dimension $p$. The inverse CDF of the normal and the logistic distributions are known as the probit and logit links respectively. In case of continuous or categorical items with the probit link, the marginal distribution of ${\bf y}^*_i$ is:
\begin{equation}
\label{gbsem_marginal}
{\bf y}^*_i\sim N(\alpha, \Lambda \Phi \Lambda^T+\Psi).
\end{equation}
However, such an expression is not available for the logit model.

The model defined by \eqref{gbsem_augmented} and \eqref{gbsem_augmented2} applies to confirmatory factor analysis (CFA) and exploratory factor analysis (EFA), and the differences between them are expressed in terms of restrictions on the parameters $\Lambda$ and $\Phi$. 
CFA postulates certain relationships among the observed and the latent variables assuming a pre-specified pattern for the model parameters (factor loadings, residual variances). For example, this is achieved by setting several elements of $\Lambda$ to zero that are referred to as cross-loadings.
EFA analyses a set of correlated observed variables without knowing in advance either the number of factors that are required to explain their interrelationships or their meaning or labelling. Finally,  SEM develops from CFA by studying the relationships between the latent variables.
 An assumption, which is common to EFA and CFA, is the conditional independence of the variables given the factors. This is equivalent to setting the off-diagonal terms in the covariance of the $\bfepsilon$, also known as error correlations, to zero. 

\subsection{Generalised Bayesian model framework}
The Bayesian SEM approach introduced in \citeA{MA12} (approximate zero framework),  covers continuous items, and it can be extended to binary and ordinal data under the probit specification.
We propose a different model specification that allows the modelling of categorical data with different links, such as the logit, but also the estimation of item-individual residuals. We extend model \eqref{gbsem_augmented} by splitting $\bfepsilon_i$ into ${\bf u}_i$, a $p$-dimensional vector of random effects with a non-diagonal covariance matrix $\Omega$, and $\bfe_i$, an error term with a diagonal covariance matrix $\Psi$. The model is:
\begin{equation}
\label{gbsem_newmodel}
{\bf y}_i^*=\alpha+\Lambda{\bf z}_{i}+{\bf u}_{i} +\bfe_i,
\end{equation}
 The item-individual specific random effect, ${\bfu}_i$,  aims to capture associations of relatively small magnitude amongst the variables, beyond those explained by the vector of latent variables $\bfz_i$. Those associations can be due to question wording, method effect, etc. Moreover, the cross loadings $\Lambda$ in \eqref{gbsem_newmodel} and \eqref{gbsem_newmarginal} are non-zero parameters that are assigned informative priors centred around zero, e.g. $N(0,0.01)$. 
For continuous normally distributed data, Model \eqref{gbsem_newmodel} coincides with the model proposed in \citeA{MA12}, written as 
\begin{equation}
\label{gbsem_newmarginal}
{\bf y}^*_i \sim N(\alpha, \Lambda \Phi \Lambda^T+\Omega+\Psi),\;\;i=1,\ldots,n,
\end{equation}
where ${\bf u}_i\sim N(0,\Omega)$, ${\bf z}_i\sim N(0,\Phi)$ and ${\bfe}_i\sim N(0,\Psi)$. Compared to the model in \eqref{gbsem_marginal}, the aim is to go from a diagonal matrix $\Psi$ to an almost diagonal one, $\Psi+\Omega$, so that the error correlations are not substantial. This can be achieved by assigning an informative prior on the non-diagonal $\Omega$ to ensure that its magnitude is low compared to $\Psi$.  

The generalised framework of \eqref{gbsem_newmodel} provides several extensions. It is  possible to define the approximate zero model for logistic models by assuming $e_{ij} \sim \text{Logistic}(0,\pi^2/3)$ (see Section \ref{gbsem_sec:logistic} for details). Other distributions (e.g. $t$-distribution, non-Normal) can also be assumed for ${\bf e}_{i}$ and ${\bf u}_i$s. Setting $\Phi=I_k$ in \eqref{gbsem_newmodel} leads to the EFA model, nevertheless fitting such a model with MCMC may challenging as we discuss later on; see also \cite{LW04,EC17,FL18,CFHP14} for some relevant Bayesian EFA schemes. 

Inference is carried by adopting a fully Bayesian framework. This requires assigning priors on all the model parameters $\theta$, denoted by $\pi(\theta)$, and proceeding based on their posterior given the data $\bfY=\{\bfy_i\}_{i=1}^n$, denoted by $\pi(\theta|\bfY)$, obtained via the Bayes theorem. A key feature of the approximate zero framework is that the priors on the cross loadings given in $\Lambda$ and the error covariances of $\Omega$ are informative and point towards zero. Next, we discuss in detail the model and prior specifications for continuous and categorical data. 

\subsubsection{Models and priors for continuous normally distributed data}
The model in \eqref{gbsem_newmodel} originates from the specification below:
 \begin{equation}
 \label{gbsem_contaug}
 \begin{cases}
{\bf y}_{i}= \alpha +\Lambda {\bf z}_i +{\bf u}_i +\bfe_i \\
{\bf z}_i \sim N(0,\Phi)\\
\bfe_i \sim N(0,\Psi)\\
{\bf u}_i \sim N(0,\Omega)
\end{cases}
\end{equation}
Nevertheless, as mentioned earlier, non-Normal distributions can be assigned on ${\bfe}_i$s, ${\bf u}_i$s and even ${\bf z}_i$. In the case where all these are assumed to be Normal, the following augmentation is also equivalent:
\begin{equation}
\label{gbsem_contmarg}
 \begin{cases}
{\bf y}_{i}\mid {\bf u}_i \sim N\big(\alpha +{\bf u}_i,\,\Lambda \Phi \Lambda^T +\Psi\big) \\
{\bf u}_i \sim N(0,\Omega).\\
\end{cases}
\end{equation}

Regarding priors, we begin with $\Omega$, the non-diagonal covariance matrix that introduces the error correlations. We use the Inverse Wishart distribution with identity scale matrix and $p+6$ degrees of freedom to reflect prior beliefs of near zero residual covariances, as is done in \citeA{MA12}. The matrix $\Phi$ is set to be a covariance matrix and primary loadings in the $\Lambda$ matrix are set to $1$ for identification purposes. 

The variances of the Normal priors assigned on the elements of $\Lambda$ depend on whether these are regarded as cross-loadings or free parameters according to the hypothesised model. The cross loadings are assigned Normal distributions with zero mean and a variance of $0.01$ as in \citeA{MA12}, whereas the remaining parameters of $\Lambda$ require some extra attention. A frequently used option is to assign large variance Normal priors, but this can lead to issues such as Lindley's paradox \cite{L57}. One way to guard against such problems is to use unit information priors \cite{KW96}. The main idea behind unit information priors is to avoid the very large prior variances causing the paradox, by setting them so that they correspond to information from a single observation point. \citeA{LW04} and \citeA{GD09}, in the context of EFA, recommend the following unit information priors: 
\begin{equation}
\label{gbsem_unit prior}
  \Lambda_{ij}\sim N(0,\psi_j^2)
\end{equation}
where $\psi_j^2$ are the idiosyncratic variances of the diagonal matrix $\Psi$ that are treated as unknown parameters. Note, however, that the above priors may cause problems in cases where the $\psi_j^2$s are quite small as it is essential to differentiate from the prior variance of $0.01$ used for the cross loadings. For this reason, a fixed value may be used instead for the prior variance of the free elements of $\Lambda$, based on preliminary estimates of them.

Regarding the diagonal matrix $\Psi$, independent Inverse Gamma priors, introduced in \citeA{FL18} and used in \shortciteA{CFHP14}, can be assigned on each $\psi_j^2$. The hyper-parameters of these Inverse Gamma priors are set in a way so that Heywood cases are given very small prior weight. More specifically, the prior given to the idiosyncratic variance is 
$$ 
\psi_j^2 \sim \text{InvGamma}(c_0, (c_0-1)/(S_y^{-1})_{jj}),
$$
where $S_y$ is the empirical covariance matrix and $c_0$ is a constant that the researcher can choose in order to limit the probability of running into Heywood issues that arise when 
$$ 
1/\psi_j^2 \geq (S_y^{-1})_{jj}. 
$$ 
Following \citeA{FL18} and \citeA{CFHP14}, the constant $c_0$ can be chosen such that the prior probability of the event above is quite small. In the application considered here, the value of $c_0=2.5$ was chosen on that basis. This is a data-dependent prior but the impact incorporates a minimal amount of information and it also helps avoid identification and MCMC convergence issues that are associated with Heywood problems. To confirm this we also conducted a sensitivity analysis that is presented in appendix A.2. The results using the chosen data-dependent prior were practically identical with those obtained using several data-independent priors.

Finally, large variance Normal priors are assigned on the $\alpha$ parameters. In every analysis that follows we use the following wide prior Normal, $\alpha \sim N(0,10^2)$. 

\subsubsection{Models and priors for binary and ordinal data}
\label{gbsem_sec:logistic}
The model for binary data using the underlying variables $y_{ij}^*, (j=1\,\ldots,p)$ is: 
\begin{equation}\label{gbsem_underlying}
\nonumber \begin{cases}
y_{ij}= \mathcal{I}(y_{ij}^*>0),\\
{\bf y}_i^* = \alpha+ \Lambda {\bf z}_i +{\bf u}_i+ \bfe_{i},\\
\bfe_ i\sim \prod_{j=1}^p \text{Logistic}(0,\pi^2/3)\;\;\text{or}\;\;\prod_{j=1}^p N(0,1)\\
{\bf z}_i \sim N(0,\Phi)\\
{\bf u}_i \sim N(0,\Omega).
\end{cases}
\end{equation}
In the above models the $\bfe_i$s correspond to the logistic and probit specifications that are the most frequently used models, although other choices of distributions are also possible. The above expressions may be simplified by integrating out the $\bfe_i$s and obtain 
\begin{equation}
\label{gbsem_binaug}
 \begin{cases}
{\bf y}_{i}\sim \prod_{j=1}^p\text{Bernoulli}\big(\pi_{ij}(\eta_{ij})\big)\\
\pi_{ij}(\eta_{ij})=\sigma(\eta_{ij}) \;\;\text{or}\;\; \pi_{ij}(\eta_{ij})=\Phi(\eta_{ij}),\;\;\eta_{ij}=[\bfeta_i]_j\\
{\bfeta}_i \coloneqq \alpha+\Lambda {\bf z}_i +{\bf u}_i,\\
{\bf z}_i \sim N(0,\Phi)\\
{\bf u}_i \sim N(0,\Omega)
\end{cases}
\end{equation}
where $\sigma(\cdot)$ denotes the sigmoid function and leads to the logit model, whereas $\Phi(\cdot)$ denotes the cumulative density function of the standard Normal distribution and leads to the probit model. Note that the distribution of ${\bf u}_i$s, and even ${\bf z}_i$s, need not be Normal under the framework, this was only done for exposition purposes. In the cases where ${\bf u}_i$s are indeed assumed to be Normal, the amount of data augmentation can be reduced further by the following equivalent formulation
\begin{equation}
\label{gbsem_binmarg}
 \begin{cases}
{\bf y}_{i}\sim \prod_{j=1}^p\text{Bernoulli}\big(\pi_{ij}(\eta_{ij})\big)\\
\pi_{ij}(\eta_{ij})=\sigma(\eta_{ij}) \;\;\text{or}\;\; \pi_{ij}(\eta_{ij})=\Phi(\eta_{ij})\\
{\bfeta}_i \sim N(\alpha,\,\Lambda \Phi \Lambda^T + \Omega).
\end{cases}
\end{equation}
In the simulation experiment and real-world examples the formulations of \eqref{gbsem_binaug} and \eqref{gbsem_binmarg} were used as they are more convenient in the context of MCMC for models based on the logit link. 

In terms of interpretation, it is interesting to note that the proposed model extends the two-parameter logistic IRT model by allowing for an item-individual random effect in addition to the standard individual latent variable $\bfz_i$. The probability of a correct response to item $j$ by individual $i$ can be written as
\bea
\frac{1}{1+\exp\left(-\left[\alpha+\Lambda{\bf z}_{i}\right]_{j}-u_{ij}\right)}.\nonumber
\eea
Similarly to the binary case, to model an ordinal observed variable $y_j$ with $m_j$ categories, we assume the existence of an underlying continuous variable $y^*_j$ so that $y_{j}= a $ if $ \tau_{a-1}^{(j)} < y^*_{j} \leq \tau_{a}^{(j)}, a=1,\ldots,m_j$.

The multinomial model is assumed to be:
$$ y_{ij} \sim \prod_{s=1}^{m_j}\pi_{j,s}(\bfeta)^{y_{j,s}}$$ where $y_{j,s}=1$ if the response $y_{ij}$ is in category $s$ and 0 otherwise, 
$\pi_{ij,s}(\eta_{ij})=(\gamma_{ij,s}(\eta_{ij})-\gamma_{ij,s-1}(\eta_{ij}))$ and $\gamma_{ij,s}(\eta_{ij})$ is a cumulative probability of a response in category $s$ or lower to item $y_j$. 
Furthermore, 
\begin{equation}
\label{gbsem_ordaug1}
 \begin{cases}
{\bf y}_i\mid \bfeta_i \sim \prod_{j=1}^p \text{Multinomial}(\pi_{ij}( \eta_{ij})) \\
\gamma_{ij}(\eta_{ij})=\sigma(\eta_{ij}) \;\;\text{or}\;\; \gamma_{ij}(\eta_{ij})=\Phi(\eta_{ij})\\
{\bfeta}_i = \tau+\Lambda {\bf z}_i +{\bf u}_i,\\
{\bf z}_i \sim N(0,\Phi)\\
{\bf u}_i \sim N(0,\Omega).
\end{cases}
\end{equation}
The parameters ${\tau}$ are unknown parameters also referred to as `cut-points' on the logistic, probit or other scale, where $\tau_{0}^{(j)}=\infty,\tau_{1}^{(j)}<\tau_{2}^{(j)}<\ldots<t_{m_j-1}^{(j)}, \tau_{m_j}^{(j)}=+\infty$. 

Similar priors can be assigned as in the case of continuous data. Regarding the elements of the $\Lambda$ matrix that are not approximate zero, unit information priors can be used. In the case of the 2PL IRT model this translates to a $N(0,4)$ prior \cite{VNM14}.

\subsection{Overview of the models and their estimation}
\label{gbsem_sec:ModelsEstimation}

In this Section, we focus on four models, within the framework defined so far, that are essential for the model assessment methodology developed in this paper. We then provide details and discussion regarding their implementation. The four models are defined below:

\begin{itemize}
  \item The exact zero (EZ) model. This is the standard structural equation model and provides the starting point in the analysis considered here. It is defined by equations \eqref{gbsem_augmented} and \eqref{gbsem_augmented2} with the cross-loadings in $\Lambda$ fixed to zero.
  \item The approximate zero (AZ) model. This is the model first introduced in \citeA{MA12} and extended here with the introduction of an item-individual random effect in the linear predictor that facilitates the modelling of categorical data with link functions different to probit as well as the detection of outliers (see Model \eqref{gbsem_newmodel}). In the case of normally distributed ${\bf u}_i$s, $\bfe_i$s and ${\bf z}_i$s, is simplified to \eqref{gbsem_newmarginal}. An important feature is that the cross-loadings in $\Lambda$ are no longer being fixed to zero. It is a model to be used only in the Bayesian sense, as the informative priors on the $\Omega$ and on the cross-loadings in $\Lambda$ are essential. 
  \item The exploratory factor analysis model (EFA). It is the standard EFA model, defined here by equations \eqref{gbsem_augmented} and \eqref{gbsem_augmented2} where low informative priors are assigned to all the components of $\Lambda$ and $\Phi=I$. 
  \item The EFA model with item-individual random effects (EFA-C). It is defined as the EFA model but with equation \eqref{gbsem_newmodel} instead of \eqref{gbsem_augmented2}. This approach to EFA can allow for a small amount of item dependencies conditional on the extracted independent factors. That model specification might result in greater amount of dimension reduction than EFA, since the stricter assumption of conditional independence could require a model with additional factors. 
\end{itemize}

In terms of implementation, it is generally possible to use MCMC and several schemes can be used, \cite<see e.g.>{E10}. In cases where the $\bfe_i$s, ${\bf u}_i$s and ${\bf z}_i$s are all assumed to be Normal, Gibbs samplers may be formed, \cite<see e.g.>{GZ96,CG98}. A Gibbs sampler may also be available in more general models using the P\'{o}lya-Gamma augmentation suggested in \shortciteA{Polson13}; see also \shortciteA{Jiang19} and \shortciteA{AM21b}.  In this paper, we recommend the use of Hamiltonian Monte Carlo (HMC) \cite{N11}, as it does not require the additional P\'{o}lya-Gamma augmentation, thus resulting in models of lower dimensionality, and also  covers all cases in terms of both models and priors. Moreover, HMC can be implemented with the help of programming frameworks such as Stan \shortcite{stan} that are well suited for assessing convergence properties of MCMC schemes, which is an essential task particularly for the models of this paper. It is also supported by high-level software packages such as `blavaan` \cite{MR15}. For a complete repository of the code and further implementation details we refer the interested reader to the code repository for this work hosted on github at `bayes-sem'\footnote{https://github.com/bayesways/bayes-sem/}. Fitting the EZ model in Stan is generally straightforward although we note that it may be useful to consider different parametrisations to improve MCMC performance and stability. For example, one may set one loading of each factor in $\Lambda$ to one and consider a full covariance matrix $\Phi$ or just restrict the leading cross loadings to be positive and consider a correlation matrix for $\Phi$. 


In cases where the EZ model does not perform well it is essential to find an appropriate benchmark to assess the performance of the AZ model. As discussed in more detail in the next section, such benchmarks can be provided by the EFA and EFA-C models. In general, fitting EFA models using MCMC can be a challenging task, due to issues such as rotational indeterminacy. The problem lies in the fact that the likelihood is specified in terms of $\Lambda\Lambda^{T}$ but often interest lies instead on $\Lambda$. The lower triangular set of restrictions \cite<see e.g.>{GZ96} ensures the mapping between those matrices is well defined, but introduces order dependencies among the observed variables. The choice of the first $k$ variables, which is an important modelling decision \cite{CCLNWW08}, thus becomes influential. The schemes of \shortciteA{CFHP14,FL18,BD11} provide an alternative to setting these restrictions and can also be used to identify the number of factors in a single MCMC run. However, as also noted in \shortciteA{BD11}, for a number of tasks such as choosing the number of factors or assessing the predictive performance, there is no need to focus on $\Lambda$, instead one can monitor $\Lambda\Lambda^{T}$ which is free of rotational issues. In such cases, the restrictions on $\Lambda$ can be omitted as long as there are no MCMC convergence and mixing issues on the $\Lambda\Lambda^{T}$ elements. As described in the next section, EFA and EFA-C models are only used in this paper to establish a benchmark for their predictive performance, hence focusing on $\Lambda\Lambda^{T}$ is sufficient. 

\section{Model assessment}
\label{gbsem_sec:assessment}
In this section, we introduce a model assessment framework that collectively uses fit indices and cross-validation to detect overfit. The aim is to complement PPP values, or other similar indices, with scoring rules to evaluate the prediction extracted from the model. The aim is to achieve a good fit and avoid overfit. The suggested procedure involves calculating these metrics for the EZ and AZ models as well as the EFA and EFA-C models with the same number of factors. We begin by presenting the proposed indices in detail, and finally provide our suggested procedure along with some guidelines and recommendations.

\subsection{Assessing goodness of fit with PPP values}
\label{gbsem_sec:ppps}

PPP values are perhaps the most frequently used method to assess model fit in the Bayesian SEM framework. Posterior predictive checking relies on a discrepancy function denoted by $D(\bfY,\theta)$ that quantifies how far the fitted model is from the data. For continuous data, the discrepancy function used here is the likelihood ratio test (LRT) function \cite<see e.g.>{SHB99} comparing the estimated model ($H_0$ hypothesis), and the unconstrained variance-covariance matrix model ($H_1$ hypothesis). The unconstrained model is also known as the saturated model (perfect fit). $D(\bfY,\theta)$ is given by:
\begin{equation}
\label{gbsem_LRT}
  \mbox{LR}\left[S,\Sigma(\theta)\right]=(n-1)\left\{\log \left|\Sigma(\theta)\right| +\mbox{tr}\left[S\Sigma^{-1}(\theta)\right] -\log\left|S\right|-p\right\},
\end{equation}

where $S$ and $\Sigma(\theta)$ are the sample and model implied variance-covariance matrix respectively. Furthermore, $\left|\cdot\right|$, $\mbox{tr}(\cdot)$ denote the determinant and trace of a matrix respectively. For example, if the maximum likelihood estimate (MLE) of $\theta$, is plugged in \eqref{gbsem_LRT}, then $\mbox{LR}\left[\cdot\right]$ is a statistic, but if $\theta$ is unknown then $\mbox{LR}\left[\cdot\right]$ may be viewed as a metric. Given the discrepancy function $D(\bfY, \theta_m)$ defined in \eqref{gbsem_LRT}, a suitable MCMC algorithm and $M$ posterior draws, the PPP value is computed as follows: 
\begin{enumerate}
\item At each (or some) of the MCMC samples $\theta_m$, $m=1,\dots,M$, do the following:

\begin{enumerate}
\item Compute $D(\bfY, \theta_m)$.
\item Draw $\bf\Ytil$ having the same size as $\bfY$, from the likelihood function $f(\bfY|\theta_m)$ of the implied model in Equation \eqref{gbsem_marginal} or \eqref{gbsem_newmarginal} and using the current value $\theta_m$. 
\item Calculate $D({\bf\Ytil}, \theta_m)$ and $d_m= \mathcal{I}\big[ D(\bfY, \theta_m)< D({\bf\Ytil}, \theta_m)\big]$, where $ \mathcal{I}\big[\cdot\big]$ is an indicator function.
\end{enumerate}
\item Return PPP$=\frac{1}{M}\sum_{m=1}^M d_m$.
\end{enumerate}

In the case of binary and ordinal data, the model is written as the probability of a response pattern. For $p$ binary items, there are $R=2^p$ possible response patterns, denoted by $\{ {\bf{y}}_r \}_{r=1}^R$, with corresponding observed frequencies denoted by $O_r$ where $r=1,\ldots,R$. The probability of a response pattern, based on the logistic model with a parameter vector $\theta$, and the assumption of conditional independence given $\bf z$ and $\bf u$ is:
\beq
\label{gbsem_eq:expected}
\pi_r(\theta)=\int \prod_{j=1}^p \text{Bernoulli}\left\{[{\bf{y}}_r]_j | \sigma([\bfeta]_j)\right\}f({\bf z})f({\bf u})d{\bf z} d {\bf u},
\eeq
where Bernoulli$\left\{y| \pi\right\}$ denotes the Bernoulli probability mass function for a binary observation $\bfy$ and probability of success $\pi$, $\bfeta$ is as defined in Section 2.2.2, i.e. $\bfeta=\alpha+\Lambda {\bf z} +{\bf u}$, and ${\bf z}$ and ${\bf u}$ are the latent components in the implied model. The integral in \eqref{gbsem_eq:expected} can be approximated using Monte Carlo. Similar expressions can also be obtained for the probit specification. An equivalent model can now be defined for the observed frequencies $\left(O_1,\dots,O_{R}\right)$ given the model-based $\pi_r(\theta)$s via the Multinomial distribution
\begin{equation}
\label{gbsem_eq:multsem}
  \left(O_1,\dots,O_{R}\right) \sim \text{Multinomial}\left[n,\pi_{1}(\theta),\dots,\pi_{R}(\theta)\right].
\end{equation}
In the context of PPP values, a frequently used discrepancy measure, \cite<see e.g.>{S05}, is the $G^2$ statistic given by
\beq
\label{gbsem_eq:g2}
D(\bfY,\theta)=\sum_{r=1}^R O_{r}\log\left(\frac{O_{r}}{n\pi_{r}(\theta)}\right).
\eeq
For a given $\theta$, e.g. a sample draw from the posterior, \eqref{gbsem_eq:g2} can be derived from the likelihood ratio between the model in \eqref{gbsem_eq:multsem} and the saturated version of it where each $\pi_{r}(\theta)$ is a separate unknown parameter that can be estimated by $O_{r}/n$.
Given $M$ MCMC samples from the posterior, the PPP value is then calculated following the steps given above for continuous data. 
PPP values are not $p$-values and therefore are not necessarily connected with the relevant type I error argument. Instead, they are regarded merely as fit indices. In terms of criteria on the PPP values, we follow the relevant discussion in \shortciteA{MA12}. As such, the fit of a model with a PPP value around $0.5$ is regarded as excellent. It is generally not clear how low a PPP value should be to warrant poor fit but usually this threshold is set to $0.1$ or $0.05$. 

The discrepancy function used here checks the overall fit of the model. Other discrepancy functions can be used that check the fit on lower order margins. In the case of categorical data, one can compute chi-square type residuals \cite<see e.g.>{joreskog.moustaki:01} on the univariate, bivariate and trivariate margins as well as utilise the work on limited information test statistics such as the $M_2$ test statistic by \citeA{maydeu.joe:05}. Those PPP values can be useful in detecting model misfit in pair or triple of items. Those discrepancies can be investigated in future research within the paper's proposed framework. 

\subsection{Scoring rules in SEM via cross-validation}
\label{gbsem_sec:logscores}

As mentioned in the introduction, it is essential to assess the out-of-sample predictive performance of each model considered in addition to its fit. Although prediction is not necessarily the main aim of factor analysis models, certain ideas from predictive inference can be borrowed here to help us assess model fit and overfit. The focus is on a model's ability to predict new data that was not used for estimating the model parameters. Hence, we divide individuals into two samples: i.) the training sample $\bfY^{tr}$ used to estimate the model parameters, through the posterior distribution $\pi(\theta|\bfY^{tr})$, and ii.) the test sample $\bfY^{te}$ used to check the forecasts of the model estimated above. 

More specifically, the predictions for the unseen data come in the form of a distribution $h(\bfY^{te}|\bfY^{tr})$ that can be contrasted as a whole against the actual test data $\bfY^{te}$. In the frequentist case, one option for such a predictive distribution is $f(\bfY^{te}|\hat{\theta}^{tr})$, where $f(\cdot)$ denotes the likelihood function and $\hat{\theta}^{tr}$ is the MLE obtained from $\bfY^{tr}$. Under the Bayesian framework, the standard choice is the posterior predictive distribution 
\begin{equation}
\label{gbsem_eq:predictive}
f(\bfY^{te}|\bfY^{tr})=\int f(\bfY^{te}|\theta)\pi(\theta|\bfY^{tr})d\theta.
\end{equation}

In order to assess the quality of these distributions, scoring rules can be used, \cite<e.g. see>{DM14, GR07} as indices whose small values typically indicate good performance. For example, one common choice for a scoring rule is the log score. For a predictive distribution $h(\bfY^{te})$ the log score is defined as
\begin{equation}
\label{gbsem_eq:lsrule}
LS(\bfY^{te}) = - \log h(\bfY^{te}).
\end{equation}
The log score is among a class of scoring rules with the desired property of being strictly proper. Strict propriety for a scoring rule ensures that the optimal model among the ones considered will be uniquely identified. More specifically, the score of this optimal model will be strictly lower than the scores of the other models; in the case of it being smaller or equal we get a proper scoring rule rather than strictly proper. 

The log score may be seen as a natural extension to the goodness of fit criterion based on the likelihood ratio test statistic for prediction assessment. Consider an SEM model, defined by \eqref{gbsem_augmented} or \eqref{gbsem_newmodel} and \eqref{gbsem_augmented2}, and suppose we want to compare it against the saturated model, e.g. in the case of continuous data the model $\bfY^{te}\sim N(\alpha, \Sigma)$ for an unconstrained variance-covariance matrix $\Sigma$. Denoting with $f^{SEM}(\cdot)$ and $f^{S}(\cdot)$ the density functions of the SEM and saturated models respectively, the difference between the two log scores becomes
\begin{equation}
  \label{gbsem_eq:LRSR}
  -\log \left[\frac{f^{SEM}(\bfY^{te}|\hat{\theta}^{tr})}{f^{S}(\bfY^{te}|\hat{\alpha}^{tr},\hat{\Sigma}^{tr})}\right].
\end{equation}
The above may be viewed as the likelihood ratio test statistic based on point parameter estimates from the training data $\bfY^{tr}$, but evaluated on the unseen test data $\bfY^{te}$. 

Note that in \eqref{gbsem_eq:LRSR}, the predictive distributions do not account for the uncertainty in the parameter estimates, which can be substantial for small training sample sizes. The Bayesian framework accounts for this source of uncertainty in a natural way via the posterior predictive distribution \eqref{gbsem_eq:predictive}.

Quite often, and in most of the Bayesian models considered in this paper, the log score is not available in closed form since the same holds for the posterior. Instead, samples from the predictive distribution are available that allow an approximation of it. A standard approach is the mixtures-of-parameters (MP); see, for example, \shortciteA{KLTG21} and in particular Table 1 of its supplementary material for a list of papers where this approximation was used. In general the MP approximation for the test data point $Y^{te}_i$ uses the following Monte Carlo approximation on the conditional predictive density $f(Y^{te}_i|\Theta)$, which is required in closed form, given samples ${\Theta}_{m=1}^M$ from the posterior $\pi(\Theta|\bfY^{tr})$
\begin{equation}
LS(Y^{te}_i) = - \log \int f(Y^{te}_i|\theta)\pi(\theta|\bfY^{tr})d\theta\approx -\log \left\{\frac{1}{M}\sum_{m=1}^Mf(Y^{te}_i|\theta_m)\right\}.
\end{equation}

For continuous data, $f(Y^{te}_i|\theta_m)$ is typically a Normal probability density function, as illustrated in Section \ref{gbsem_sec:models}; see for example equations \eqref{gbsem_marginal} or \eqref{gbsem_newmarginal}. 

For the cases of binary or ordinal data, it is possible to compute the log score via the alternative formulation based on frequency patterns. Note that the posterior predictive density is given by Equations \eqref{gbsem_eq:multsem} and \eqref{gbsem_eq:expected} where the integral in the latter is with respect to the posterior based on the training data. We can therefore write the log scoring rule for a set of observed frequencies in the test data $\mathbf{O}^{te}=\left(O_1^{te},\dots,O_{R}^{te}\right)$ based on probabilities $\mathbf{\pi}^{tr}=\left(\pi_{1}(\theta)^{tr},\dots,\pi_{R}(\theta)^{tr}\right)$, obtained based on the posterior from the training data, as
\begin{equation}
\label{gbsem_eq:LS}
  LS(\mathbf{O}^{te})=-\log f(\mathbf{O}^{te}|\mathbf{\pi}^{tr})
  =-\log\left[c\prod_{r=1}^R \left[\pi_r(\theta)^{tr}\right]^{O_{r}^{te}} \right]=-\sum_{r=1}^R
O_{r}^{te}\log \pi_{r}(\theta)^{tr}+c,
\end{equation}
where $c$ represents a constant. Note that the log score only differs as a metric to $G^2$ by a constant, which essentially confirms the argument made earlier in Equation \eqref{gbsem_eq:LRSR}, about the connection of the likelihood ratio test and the log score, and makes it more specific to categorical data. 

So far we have assumed a single split between the training and test data. In order to limit the effect of peculiar splits, cross validation may be used. It is  also interesting to note that the calculation of both PPP values and scoring rules are based on the posterior predictive distribution. Nevertheless, there is an essential difference between the two approaches. PPP values are based on the posterior distribution conditional on the entire dataset and the prediction is made again on the entire dataset. In the scoring rules approach the posterior is conditional only on a subset of the data (training sample) and the prediction is made on the complement of that set (test sample).

\subsection{Model assessment with fit and predictive performance indices}
\label{gbsem:sec_recommendations}

Our procedure contains two main elements: assessing goodness of fit, as done routinely under current practice, but also assessing out-of-sample predictive performance. For goodness of fit, the governing well-known procedure is to check if the fit of the hypothesised model, the EZ model in our framework, is no worse than that of the unconstrained model (also known as the saturated model). As described earlier this can be checked by looking at the PPP values of the EZ model. In case of satisfactory PPP value, our recommendation is no different than the standard course of action, to support the hypothesised model, and there is no need to look further.

Now let us consider a situation where the EZ model does not fit the data well, in terms of a PPP value, but the AZ model does. One of the main arguments of this paper is that the researcher should not rush to support the hypothesised model as the satisfactory PPP value may as well be due to the AZ model overfitting the data. Our definition of `overfit', specifically to the SEM context is the following: If the AZ model is better than its EZ counterpart in terms of goodness of fit but also worse in terms of out-of-sample predictive performance, then it overfits the data. In other words, if the gains in goodness of fit of the AZ model, over the corresponding EZ model, are not based on systematic patterns of the data, then these gains would be of no help when predicting unseen data. Moreover the slightly increased model complexity of the AZ model may have an adverse effect in terms of prediction over the corresponding EZ model. In other words, we are seeking parsimony in addition to goodness of fit. Hence, according to our suggested framework, if the AZ model is worse than the EZ model in terms of the relevant scoring rule, then there is little support in the data for the hypothesised model. 

Next, let's consider the case where the AZ model has good PPP value, in contrast with the corresponding EZ model, and also better predictive performance as measured by the relevant scoring rule. Our view in this case is to conduct further checks. There is a possibility that AZ model is just improving upon a poorly specified EZ model but there may exist other EZ or AZ models that predict even better. If the poor fit of the EZ model is due to only some small cross loadings or error correlations, then the AZ model that captures these quantities model should perform really well. But if there are also some other systematic patterns missing from the EZ model, the improvement offered by the AZ model would be limited. Ultimately, the question that we want to answer is whether the predictive performance of each one of these models is good enough. Therefore, it is essential to establish a benchmark when comparing predictive performances. In the case of goodness of fit assessment this is done by the performance of the unconstrained saturated model. But this may not be a suitable choice for assessing predictive performance \cite{MRN92}. The problem lies in the fact that the saturated model has substantially higher complexity, or else substantially larger number of parameters, than the hypothesised models. Generally speaking, if two models with different numbers of parameters have similar in-sample performance, then the one with the smaller number of parameters will generally perform better out of sample. An alternative option for a benchmark model, exploited in this paper, is the EFA model with the same number of factors as the hypothesised model. This model has generally fewer parameters than the saturated one and is generally expected to perform well in terms of predictive performance as it is allowed to search for systematic patterns in the data without any restrictions, other than having $k$ factors. This is not the case for the EZ and AZ models, where explicit restrictions are given and it is often hoped that they will not be too far from those indicated from the EFA. Hence, in order to regard the predictive performance of the hypothesised model as satisfactory, its scoring rule should be comparable with that of the EFA model chosen as the benchmark.

Caution must be exercised over the choice of the benchmark EFA model, as selecting an over-parameterised EFA model will set the bar too low in terms of predictive performance. Therefore it may be more appropriate, in some cases, to check the parsimony of the EFA model selected as the benchmark. This can be done, for example, by considering EFA models with fewer factors, provided that they fit the data well. In order to select the number of factors $k$, models with different $k$s can be fit separately and compared. The comparison can be done by standard indices, such as the model evidence, BIC, etc., or via their goodness of fit and prediction ability as described earlier. 

In line with such considerations, we note that the presence of the small error correlations induced by the ${\bf u}_{i}$s under the approximate zero framework may offer an advantage to CFA models in terms of prediction as it can be viewed as an additional minor factor. Hence, in order to bring CFA and EFA models onto a level playing field, it may be reasonable to incorporate small error correlations to both of them via the EFA-C model.

Note also that, while the AZ models are more flexible than their EZ counterparts, they can still perform badly in cases of substantial model misspecification. For example, in cases of large enough cross loadings, say more than $0.5$, using the Normal$(0,0.01)$ as prior can still result in poor performance compared to the EFA model. This comparison may thus be exploited to detect misspecified models as we illustrate in Sections \ref{gbsem_sec:simulations} and \ref{gbsem_sec:realdata}.

We summarise below the recommendations of our proposed framework.

\ben
\item If the EZ model has satisfactory fit indices such as PPP values, there is strong support towards the hypothesised model.

\item If both EZ and AZ models have poor PPP values then there is little support of the hypothesised model. Perhaps it may be useful to use more vague priors to explore its weaknesses. It would be expected in this case that the EFA models will have better predictive performance otherwise there maybe issues in the fitting algorithms or elsewhere.

\item If the EZ model has poor performance in terms of fit indices, whereas the AZ model is satisfactory, it is essential to check the scoring rules. If the improvement offered by the AZ model is due to overfit, it is expected that the prediction score for the AZ model will be inferior to that of the EZ one. 

The predictive performance of models that overfit is therefore expected to diminish. On the other hand, a prediction score that still favours the AZ model suggests that overfit is not the case. To check if the predictive performance of the AZ model is good enough, comparisons with EFA type models can be made. In cases of comparable or improved performance there is supporting evidence towards the hypothesised model.
\een

Model fit assessment is by no means an easy task especially in factor analysis modelling where model misfit can be due to various reasons such as misspecification of the latent variable distribution, item dependencies, skewed data and non-linear predictors. In this paper, for the calculation of PPP values we use a discrepancy function that looks at the overall fit of the model both in the case of continuous and categorical data. 
 It is useful to complement those overall goodness of fit tests with other measures of fit such as residuals and limited information test statistics that check the fit on lower order margins and detect item misfit as explained in Section 3.1.
 It is because of those complexities that our proposed methodology is trying to shed light to model fit challenges using a different set of tools that look at the model's out of sample prediction performance. This provides new tools within the Bayesian modelling framework in SEM and highlights even further the challenges of fit and problems of PPP values. 
 Furthermore, the new residual term in the linear predictor defined by the item-individual random effects ${\bf u}_i$s plays a key role since it can be used as model diagnostics to detect outliers such as leaked items and cheating behavior in educational testing or secondary response strategies employed by some of the respondents to some of the items.

\section{Simulation experiments}
\label{gbsem_sec:simulations}

\subsection{Setup}
 
Simulation experiments were conducted to study the performance of the proposed models and demonstrate the assessment framework for continuous and binary data. We focus on two cases of data generated using Equation \eqref{gbsem_augmented2}, i.e. continuous and binary. For each of these two cases, three scenarios were considered when generating simulated data:
\begin{itemize}
\item Scenario 1: Data generated from the EZ model.
\item Scenario 2: Data generated from the AZ model with small error correlations, introduced by item-individual random effects, and without cross loadings. 
\item Scenario 3: Data generated from the AZ model with two non-negligible cross loadings and without small error correlations. 
\end{itemize}

For both continuous and binary data, we considered $p=6$ items and $k=2$ factors. The factor loadings used to generate the data, in each of the three scenarios, are shown in Table \ref{gbsem_table:nzcl}. 
\begin{table}[!htbp]
\centering
\begin{tabular}{*6c}
\toprule
 \multicolumn{2}{c}{Scenario 1} & \multicolumn{2}{c}{Scenario 2} & \multicolumn{2}{c}{Scenario 3} \\
\midrule
$z_1$ & $z_2$ & $z_1$ & $z_2$ & $z_1$ & $z_2$ \\
\midrule
 1   & 0  & 1  & 0 & 1 & 0 \\
 .8  & 0  & .8 & 0 & .8 & 0 \\
 .8  & 0  & .8 & 0 & .8 & .6 \\
0   & 1  & 0  & 1 & .6 & 1 \\
0   & .8 & 0 & .8 & 0 & .8 \\
0   & .8 & 0  & .8 & 0 & .8 \\
\bottomrule
\vspace{0.1cm} \end{tabular}
\caption{True factor loadings used in the three simulation scenarios.}
\label{gbsem_table:nzcl}
\end{table}
Although the data were generated under the three scenarios, in all of them the typical hypothesised model assumes a simple structure in which the first three items load on the first factor whereas the last three load on the second factor. In other words, for the AZ model, the first three elements of the first $\Lambda$ column and the last three of the second $\Lambda$ column are regarded as the major parameters, whereas the other elements of $\Lambda$ are cross-loadings. In all three scenarios, the factor correlation was set to $0.2$, and the intercepts $\alpha$ were all zero. The sample sizes were set to $n=1,000$ in the continuous data and $n=2,000$ in the binary data. For Scenario 2, equation \eqref{gbsem_newmarginal} was used by setting the matrix $\Omega+\Psi$ to have ones in the diagonal, and $6$ non-zero off-diagonal elements set to $0.2$ with the remaining $9$ off-diagonal elements set to zero. 

For each scenario, the proposed model assessment framework of Section \ref{gbsem_sec:assessment} was put into action by computing the PPP values and scoring rules for all the previously mentioned models. After fitting and summarising these models, according to Sections \ref{gbsem_sec:ppps} and \ref{gbsem_sec:logscores}, we proceeded according to the recommendations of Section \ref{gbsem:sec_recommendations}. 

The models and priors were specified as outlined in Section \ref{gbsem_sec:models} and samples from the posterior of each model were obtained using Hamiltonian MCMC programmed using the Stan language. In the case of continuous data, $1,000$ iterations were used as the warm-up period and another $2,000$ for inference purposes. In the case of binary data, it was $2,000$ for warm-up and $2,000$ for analysis purposes. The models were run in $4$ parallel chains in each case resulting in $4 \times 2,000 = 8,000$ posterior draws. In all cases, we ensured successful convergence of the chains with the help of the automatic metrics implemented in Stan as well as visual inspection of the posterior draws. 

In all instances, we applied a 3-fold cross-validation and aggregated the scores by summing. Given that a scoring rule is a comparative index, we reported the difference in scores between each model and the best model. In other words, the best model of each case, or else the one with the smallest score, was given the value of zero.
  
The next two sections present the results of the simulation experiments for continuous and binary data. The aim of these experiments is to illustrate the performance of the proposed model framework and provide a proof of concept. More detailed simulation experiments will be helpful, as we discuss in the next sections, and are left for future research.

\subsection{Continuous data}\label{gbsem_cont_sim_description}
Table \ref{gbsem_table:cont_sim_results_diff} gives the log score ($LS$) and the PPP values for the three simulation scenarios. Starting with Scenario 1, we note that all models fit the data well in terms of the PPP values. In terms of predictive performance, we note that the $EZ$ model performs best, which is not surprising given that the data were generated from it. Note that the EZ model even improves upon the EFA models in terms of predictive performance as it is a more parsimonious model. 

\begin{table}[!htbp]
\centering
\begin{tabular}{*7c}
\toprule
{} & \multicolumn{2}{c}{Scenario 1} & \multicolumn{2}{c}{Scenario 2} & \multicolumn{2}{c}{Scenario 3} \\
\midrule
Model & PPP & LS  & PPP & LS   & PPP & LS         \\
EZ   & 0.66  & 0   & 0.03  & 3.87  & 0.00  & 39.87  \\
AZ   & 0.51  & 2.42 & 0.44  & 0.06    & 0.53  & 1.37   \\
EFA   & 0.62  & 1.14 & 0.19  & 2.85  & 0.59  & 0    \\ 
EFA-C  & 0.53  & 2.49 & 0.49  & 0  & 0.56  & 1.02   \\
\bottomrule
\vspace{0.1cm} \end{tabular}
\caption{Simulation Results for Continuous Data. PPP values and sum of log scores (LS) of 3-fold cross validation for the relevant models. For each scenario, the best model has $0$ log score and the differences from it are reported for the other models.}
\label{gbsem_table:cont_sim_results_diff}
\end{table}

In simulation Scenario 2, both the EZ and EFA models exhibit poor fit according to their PPP values, which is again not surprising given that these models assume zero error correlations. In contrast, the AZ and EFA-C models that allow for small, yet not exactly, zero error correlations both fit well. At this point, the question is whether the improved fit of the AZ model is due to fitting noise or else overfit as defined in recommendation 2 of Section \ref{gbsem:sec_recommendations}. But if AZ was overfitting the data, we would not expect to see an improved performance over the EZ model, as we see here. We can expand the investigation of the AZ model further, wondering whether there is another theory that leads to an AZ model with even better predictive performance. Recommendation 3 of Section \ref{gbsem:sec_recommendations} may shed light on this question when we compare the predictive performance against the EFA models. We see that AZ is quite competitive against those models as its log scores is almost the same as that of EFA-C. Hence, according to our proposed framework, there is strong support towards the AZ model and, consequently, the hypothesised model. This appears to be a reasonable conclusion in the SEM context given that the poor fit is due to error correlations that are usually linked with observation error rather than factor loading misspecifications. The use of PPP values alone would not have been enough to reach that conclusion.

Finally, let us consider the simulations for Scenario 3, where the EZ model does not have a good fit, as one would expect, but all the other models have PPP values around $0.5$. As before, we are interested in whether the AZ model overfits and what conclusions we can draw on the hypothesised model. To answer such questions, we set the benchmark model to be the EFA model with the higher predictive performance; it is the EFA model in this case, as one would expect since the data were simulated without error correlations. The log score of the AZ model is again much better than that of the EZ, as it utilises its approximate zero cross loadings to pick up the two cross loadings of $0.6$. But it is not better than the EFA model, thus not ruling out the presence of a different hypothesised theory regarding the loading structure of the six items. Indeed, the theory corresponding to factor loadings according to Scenario 3 described in Table \ref{gbsem_table:nzcl} provides a better model as the data were simulated from it. 

\subsection{Binary data}\label{gbsem_bin_sim_description}

In this section, we summarise the results of the three simulation experiments for the case of binary data. Table \ref{gbsem_table:bin_sim_results_diff} gives the PPP values and the log scores. The results are very similar to the continuous case. In the case of Scenario 1, all models demonstrate good fit as indicated by the PPP values. Furthermore, the EZ model is the optimal one in terms of predictive performance, which is reassuring since data were simulated from it. In Scenario 2, we see that the EZ model exhibits very poor fit, caused by the additional error correlations in the simulated data, as indicated by the PPP value of 0.02. The rest of the models exhibit a moderately good fit with PPP values above 0.10. Similarly to the continuous case, the AZ model does well in terms of both recommendations 2 and 3 of Section \ref{gbsem:sec_recommendations}, being the model with the best predictive performance. Finally, in Scenario 3, in terms of model fit the EZ model also fails, due to the presence of non-zero cross loadings. The other models do well, leaving some questions open in terms of the validity of the hypothesised theory. For this reason, according to recommendation 3 of Section \ref{gbsem:sec_recommendations}, compares the predictive performance of AZ against the best performing EFA model. In this case, the AZ model is not as good as the EFA.

\begin{table}[!htbp]
\centering
\begin{tabular}{*7c}
\toprule
{} & \multicolumn{2}{c}{Scenario 1} & \multicolumn{2}{c}{Scenario 2} & \multicolumn{2}{c}{Scenario 3} \\
\midrule
Model  & PPP& LS   & PPP& LS   & PPP& LS   \\
EZ   & 0.52  & 0    & 0.02  & 4.19   & 0.00  & 7.31   \\
AZ   & 0.50  & 0.68   & 0.12  & 0    & 0.52  & 1.90   \\
EFA   & 0.59  & 1.45   & 0.13  & 0.09   & 0.45  & 0    \\
EFA-C  & 0.54  & 3.27   & 0.17  & 0.24   & 0.50  & 2.96   \\
\bottomrule
\vspace{0.1cm} \end{tabular}
\label{gbsem_table:bin_sim_results_diff}
\caption{Simulation Results for Binary Data. PPP values and sum of log scores (LS) of 3-fold cross validation for the relevant models. For each scenario, the best model has log score equal to $0$ and the differences from it are reported for the other models.}
\end{table}

\subsection{Parameter recovery for the AZ model in the binary data case.}

To investigate the parameter recovery performance of the AZ model in the binary data case, we performed a simulation experiment where $100$ different datasets were simulated and the AZ model was fitted on each one of them to obtain samples from its posterior. More specifically the data were drawn from the EZ model, so that we focus on the main parameters of interest, namely the factor loadings and the correlation of the factors, each with sample of size $2,000$. The factor loadings used to simulate the data are the same as in the Scenario 1 of the simulation experiments of Section \ref{gbsem_sec:simulations} and the correlation between the two factors was $0.2$. Finally, the intercept parameters used to simulate the data were all set to $0$. We used the parameterisation where the loadings are unrestricted and the factors' variance is fixed to $1$ hence their covariance matrix is restricted to be a correlation matrix.

Regarding the prior specification of the AZ model, we assumed that, according to the hypothesised theory, the first factor loads on the first 3 items and the second factor loads on the last 3 items. Hence, the rest of loading parameters were regarded as cross-loadings, and were assigned informative priors around zero. The rest of the priors were assigned as described earlier in the paper.

As informal measures of how well the parameters are recovered, we focused on frequentist properties of some estimators derived from the posterior samples. The estimators consisted of the $95$\% credible intervals as interval estimators, obtained from the sample $2.5$-th and $97.5$-th points extracted from the posterior draws, as well as the posterior mean and median as point estimator. We then examined the coverage probability of the former and the bias of the latter. We note that these summaries ($95$\% credible intervals, posterior mean, and posterior median) may not exhibit the desired frequentist performance even in the case the model fits the data well, as they have not been constructed to do so. Nevertheless, if they happen to perform well, it is definitely reassuring. 

We examined the main parameters of interest, such as the loadings $\Lambda$ and the factor correlation $\rho$. The results are summarised in Table \ref{gbsem_table:parameter_recov} and they contain coverage probabilities and biases of the previously mentioned posterior summaries. As we can see, the coverage probabilities are reasonably close to $0.95$ whereas the biases are not substantial, particularly for the posterior median. We therefore conclude, while noting the informal nature of the experiment, than no substantial concerns regarding parameter recovery are raised.

\bigskip

\begin{table}[!htbp]
\centering

\begin{tabular}{*5c}
\toprule
Parameter & True Value & Coverage Rate & Bias of Post. Mean & Bias of Post. Median \\
\midrule
$\Lambda_{[1,1]}$ & 1.0 & 0.94 & 0.06 & 0.03 \\
$\Lambda_{[2,1]}$ & 0.8 & 0.96 & 0.05 & 0.03 \\
$\Lambda_{[3,1]}$ & 0.8 & 0.94 & 0.05 & 0.03 \\
$\Lambda_{[4,1]}$ & 0.0 & 1.00 & 0.00 & 0.00 \\
$\Lambda_{[5,1]}$ & 0.0 & 1.00 & 0.00 & 0.00 \\
$\Lambda_{[6,1]}$ & 0.0 & 1.00 & 0.00 & 0.00 \\
$\Lambda_{[1,2]}$ & 0.0 & 1.00 & 0.00 & 0.00 \\
$\Lambda_{[2,2]}$ & 0.0 & 1.00 & -0.01 & -0.01 \\
$\Lambda_{[3,2]}$ & 0.0 & 1.00 & 0.00 & 0.00 \\
$\Lambda_{[4,2]}$ & 1.0 & 0.99 & 0.03 & 0.00 \\
$\Lambda_{[5,2]}$ & 0.8 & 0.95 & 0.06 & 0.04 \\
$\Lambda_{[6,2]}$ & 0.8 & 0.99 & 0.03 & 0.02 \\
$\rho$    & 0.2 & 1.00 & -0.01 & -0.01 \\
\bottomrule
\vspace{0.1cm} \end{tabular}
\caption{True values, $95\%$ coverage success rate and bias of point estimators out of 100 replications, AZ model for binary data.}
\label{gbsem_table:parameter_recov}
\end{table}

\section{Real-world data examples}
\label{gbsem_sec:realdata}

In this section, we demonstrate our proposed model assessment framework with two real datasets. The first dataset is a popular psychometric test, usually referred to as the `Big 5 Personality Test', that decomposes human personality along $5$ main traits using 15 items measured on a 7-point likert scale. The second data set is based on the Fagerstrom Test for Nicotine Dependence (FTND) that consists of six binary variables.

\subsection{Example 1: `Big 5 Personality Test'}

 The data were collected as part of the British Household Panel Survey in 2005-06 focusing on female subjects between the ages of 50 and 55; the sample size consists of 589 individuals. The `Big 5 Personality Test', as it is known, is a $15$-item questionnaire on topics of social behaviour and emotional state. Participants answer each item on a scale from $1-7$, $1$ being `strongly disagree' and $7$ being `strongly agree'. Items are treated here as continuous. The test is designed to measure five major, potentially correlated, personality traits. Each trait corresponds to a factor, and each factor is hypothesised to explain exactly 3 out of 15 items.

The data have been analysed in several papers including \citeA{MA12}, \citeA{SMSD15}, and \citeA{AMM15}. In these analyses, an interesting finding was that the exact zero (EZ) model did not exhibit good fit based on several standard indices including the PPP values. The approximate zero (AZ) model gave a good fit in terms of the PPP values, but also had many non-zero error correlations. This raised concerns over whether the flexibility of the AZ model is picking up noise, thus resulting in a misleadingly high PPP value. The validity of the `Big 5' scale on these data remains unclear. In an attempt to shed more light on this question we apply our model assessment framework and summarise the results in Table \ref{gbsem_table:big5res_diff}.

\begin{table}[!htbp]
\centering
\begin{tabular}{*5c}
\toprule
Model & PPP &  LS\\
\midrule
EZ   & 0.0  & 174.64    \\
AZ   & 0.23 & 0  \\
EFA   & 0.00 & 58.24  \\
EFA-C  & 0.38 & 3.49   \\
\bottomrule
\vspace{0.1cm} \end{tabular}
\caption{`Big 5' personality test data, BHPS. PPP values and sum of log scores (LS) of 3-fold cross validation for the relevant models. For each scenario, the best model has $0$ log score and the differences from it are reported for the other models.}
\label{gbsem_table:big5res_diff}
\end{table}

The picture is very similar to the error correlations scenario in Section \ref{gbsem_cont_sim_description}, yet much more pronounced. Our analysis confirms the poor fit of the EZ and the EFA with five factors. Both AZ and EFA-C models have reasonably good PPP values. This implies that error correlations contribute to the lack of fit to a large extent. In order to assess the question of overfit and draw conclusions on the validity of the `Big 5' scale, we calculate the log scores for each model. The log score of the AZ model clearly dominates all the other models, suggesting that the model is fitting consistent patterns in the data and it clearly outperforms the EFA models. This points to strong support towards the `Big 5' scale, attributing the fit issues of the EZ model to error correlations that could have been caused by the wording and other issues often present in survey data like the BHPS.

\subsection{Binary Data: Fagerstrom Test for Nicotine Dependence}

In this section we use data on 566 patients available through the National Institute on Drug Abuse (study: IDA-CTN-0051). The Fagerstrom Test for Nicotine Dependence (FTND) \shortcite{heatherton.ea:91} was designed to provide a measure of nicotine dependence related to cigarette smoking. It contains six items that evaluate the quantity of cigarette consumption, the compulsion to use, and dependence. 
The original scale consists of 4 binary and 2 ordinal items for self-declared smokers:
\begin{enumerate}
\item FNFIRST: How soon after you wake up do you smoke your first cigarette?
[`3'=Within 5 minutes, `2'=6 - 30 minutes, `1'=31 - 60 minutes, `0'=After 60 minutes]
\item FNGIVEUP: Which cigarette would you hate most to give up? [`1'=The first one in the morning, `0'=All others]
\item FNFREQ: Do you smoke more frequently during the first hours after waking than during the rest of the day? [`1'=Yes, `0'=No]
\item FNNODAY: How many cigarettes/day do you smoke? [`0'=10 or less, `1'=11-20, `2'=21-30, `3'=31 or more]
\item FNFORBDN: Do you find it difficult to refrain from smoking in places where it is forbidden (e.g., in church, at the library, in cinema, etc.)? [`1'=Yes, `0'=No]
\item FNSICK: Do you smoke if you are so ill that you are in bed most of the day? [`1'=Yes, `0'=No]. 
\end{enumerate}
For the purposes of our analysis, item FNFIRST was dichotomised as `1'=[3] and `0'=[0,1,2] and item FNNODAY as `1'=[2,3] and `0'=[0,1].

The mapping between the FTND scale and a CFA model is not clear, see e.g. \citeA{RR05} and references therein. \citeA{RR05} fitted a single factor, a correlated two factor, and a two factor model with one cross loading. These models were also considered in our analysis and are denoted as 1F, 2F-EZ, and 2F EZ-b respectively. More specifically, under the EZ model items 1, 2 and 3 load on a `morning' smoking factor, whereas items 4, 5 and 6 load on a `daytime' smoking factor. The EZ-b model is specified by letting item `FNFIRST' load on both factors. In addition to these models, we also considered their approximate zero versions, denoted as 1F-C, 2F-AZ, and 2F-AZ-b respectively, as well as the two-factor EFA models with and without error correlations (2F-EFA and 2F-EFA-C). The results are shown in Table \ref{gbsem_table:fnd_results2_diff}.

\begin{table}[!htbp]
\centering
\begin{tabular}{*3c}
\toprule
Model & PPP & LS      \\
\midrule
1F     & 0.01 & 15.98    \\
1F-C    & 0.32 & 6.63    \\
2F-EZ    & 0.04 & 10.45    \\
2F-AZ    & 0.40 & 6.23   \\
2F-EZ-b   & 0.41 & 0.00   \\
2F-AZ-b   & 0.44 & 2.01   \\
2F-EFA   & 0.44 & 2.66   \\
2F-EFA-C  & 0.58 & 2.38    \\
\bottomrule
\vspace{0.1cm} \end{tabular}
\caption{PPP values and sum of log scores (LS) of 3-fold cross validation for the relevant models. The models with `-b' refer to the measurement model with the first item loading to both factors. The best model had log score equal to $0$ and the differences from it are reported for the other models.}
\label{gbsem_table:fnd_results2_diff}
\end{table}

Examination of the PPP values reveals concerns about the fit of the models 1F and 2F-EZ, so these are ruled out of the discussion. This raises several questions: Is the 2F-EZ-b the best model or do any of the AZ model versions, 2F-AZ or 2F-AZ-b, do better? Is the best of these three good enough? Perhaps more importantly, which measurement scale should be used for the FTND test on the basis of this dataset? 
We attempt to shed light on these questions with the use of cross-validated log scores. The best model is the 2F-EZ-b correcting the misspecifications of 2F-EZ with a single additional parameter. The fact that the log score of 2F-EZ-b is smaller than that of the EFA models provides support towards the scale with two correlated factors where the item `FNFIRST' loads on both of them. 

\section{Discussion}

In this paper, we adopt the Bayesian SEM framework, introduced in \citeA{MA12} and enhance it by providing tools of model exploration and assessment that go beyond goodness-of-fit testing. Those tools incorporate scoring rules combined with cross-validation to the existing fit indices.
The model is also extended by introducing an item-individual random effect that allows modelling of categorical data and data with other distributions than the normal. Another potential use of this random effect term, is to estimate item-individual residual values to obtain model diagnostics information for the detection of two-way outliers, e.g. leaked items and cheating behaviour in educational testing or secondary response strategies employed by some of the respondents to some of the items.   

As illustrated on simulated data and real-world examples, the use of the scoring rules can prove quite useful in SEM analysis. Nevertheless, as with any index, it would be helpful to explore it further and get a better understanding of the range of values indicating a good model in different settings. This range may depend on the sample size, the number of factors and parameters, the type of the data, the choice of the scoring rules, the number of folds or the form of cross-validation in general, the choice of the benchmark model etc. Another important component, present in any form of Bayesian analysis, is the prior specification. The behaviour of the scoring rules under different priors, e.g. the spike and slab priors, as in \citeA{LCL16}, rather than the ridge-type priors, is also an interesting question. 

The calculations can be implemented using MCMC through standard user-friendly software like Stan and can be combined with existing packages for SEM. This opens up the possibility of using fast approximate methods such as Variational Bayes \shortcite{KTRGB17} that are automated and readily available. This can be particularly useful in categorical data applications where the use of MCMC and the presence of high-dimensional latent variables can result in computation times that are larger than the users' expectations. Moreover, Variational Bayes can be used to improve the efficiency of MCMC samplers. 

Further extensions of the generalised family of models can also be explored; for example, non-Normal errors $\bfe_i$s or random effects ${\bf u}_i$s. Inspection of the latter may also provide diagnostic information for detecting outliers and removing items to purify the constructs. It would also be interesting to explore the connections with Bayes factors, as they tend to provide parsimonious models that typically do well in terms of cross-validation. Calculating Bayes factors is not always straightforward and they are also more sensitive to the choice of priors. However, such issues can be alleviated by suitable choice of priors, as done in this paper. 

Finally, it is important to note that the developed model assessment framework and the CV index can be applied outside the Bayesian SEM context. In fact, it can be useful in situations where we need to assess the fit of a more flexible model, such as semi-parametric or non-parametric formulations \cite{YD10,SLCI13}. In such models, attaining a good fit is not always associated with a good systematic part of the model, as the flexibility in its error part can lead to overfitting. Such models arise in many scientific areas and go well beyond the SEM framework.

%% file: appendix.tex
\begin{appendices}

\section{Inverse Wishart}\label{gbsem_app:invwishart}
We recall here that the Inverse Wishart distribution $\IW(D_p, d)$ is parameterised by matrix $D_p$ of dimension $p \times p$ and $d$ degrees of freedom where we need $d > p+1$ for the distribution to be well defined. The higher the value of $d$ the more concentrated the distribution gets around $D_p$. For example, if we choose $D_p = I_p$ the identity matrix of size $p$, then the marginal distribution of the diagonal elements will be distributed with mean $1/(d-p-1)$ and variance $2/[(d-p-1)^2 (d-p-3)]$, whereas the off-diagonal elements will be distributed with mean $0$ and variance $1/[(d-p)(d-p-1)^2(d-p-3)]$. Note that these expressions simplify when, for example, $d$ is set to $p+6$. We refer the interested reader to the appendix of \citeA{MA12} for more information. 

\section{Sensitivity analysis for data-dependent priors}

\begin{figure}
\centering
\includegraphics[width=0.8\textwidth]{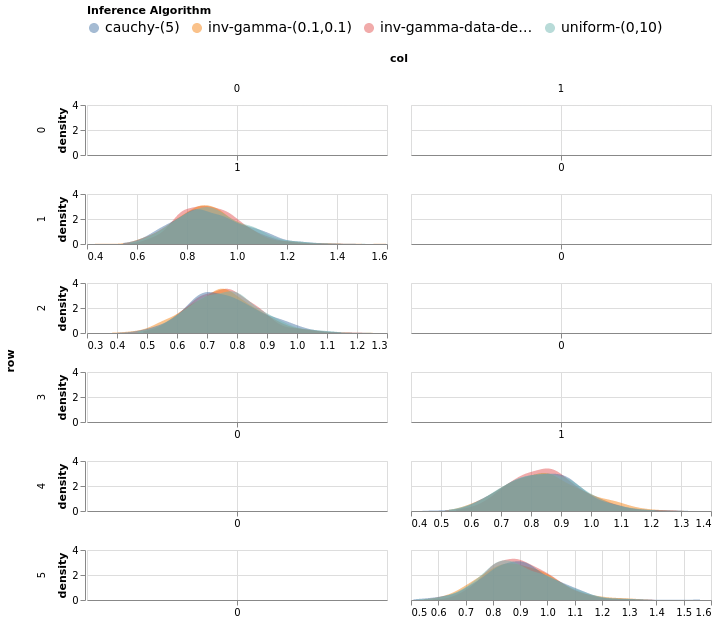}
\caption{Posterior density plots of the loading matrix parameters under 4 different prior choices. The model using a data-dependent prior (red) produces identical posterior density plots as three other models using priors independent of the data.}
\label{gbsem_fig:fig1}
\end{figure}

We performed a sensitivity analysis to examine the effect of the data-dependent priors on the final result. In order to amplify the prior effect we used a relatively small sample size, by simulating 200 data points from a standard two-factor model, according to simulation Scenario 1 in Section \ref{gbsem_sec:simulations}. We fit the EZ model with the data-dependent prior of \shortciteA{FL18,CFHP14} that protects against Heywood cases for the idiosyncratic variances $$ \psi_j^2 \sim \text{InvGamma}(c_0, (c_0-1)/(S_y^{-1})_{jj})$$ with $c_0 = 2.5$. Moreover, the following data-independent priors were also used: $\text{InvGamma}(0.1, 0.1)$, $\text{Half-Cauchy}(5)$, and $\text{Uniform}(0,10)$. The posterior samples from all four priors were used to produce kernel density plots for the posterior of the free $\Lambda$ elements that are depicted in Figure \ref{gbsem_fig:fig1}. As we can see, the posterior density plots are almost identical for all these priors. Similar results were also obtained for the remaining parameters. We therefore conclude that the data-dependent prior does not impact the final results, while helping guard against Heywood cases.

\end{appendices}